\documentclass[]{aa}
\usepackage{natbib}
\usepackage{graphicx}
\usepackage[varg]{txfonts}
\usepackage[colorlinks=true,citecolor=blue]{hyperref}
\begin{document}

   \title{Can the long-term hemispheric asymmetry of solar activity result from fluctuations in dynamo parameters?}

   \author{Alexander Nepomnyashchikh\inst{1}
          \and
          Sudip Mandal\inst{2}
          \and
          Dipankar Banerjee\inst{2,3}
          \and
          Leonid Kitchatinov\inst{1}
          }

   \institute{Institute for Solar-Terrestrial Physics, Lermontov Str. 126A, Irkutsk 664033, Russia\\
              \email{kit@iszf.irk.ru}
        \and
            Indian Institute of Astrophysics, Karamangala, Bangalore 560034, India
        \and
            Center of Excellence in Space Sciences India, IISER Kolkata, Mohanpur 741246, West Bengal, India
             }

   \date{Received DD Month Year/accepted DD Month Year}
  \abstract
{The hemispheric asymmetry of sunspot activity observed possesses a regular component varying on a time scale of several solar cycles whose origin and properties are currently debated.}
{This paper addresses the question of whether the long-term hemispheric asymmetry can result from random variations of solar dynamo parameters in time and latitude.}
{Scatter in the observed tilt angles of sunspot groups is estimated to infer constraints on fluctuations in the dynamo mechanism for poloidal field regeneration. A dynamo model with fluctuations in the Babcock-Leighton type $\alpha$-effect is designed in accordance with these constraints and then used to compute a large number of magnetic cycles for statistical analyses of their hemispheric asymmetry.}
{Hemispheric asymmetry in the simulated dynamo results from the presence of an equator-symmetric part in the oscillating magnetic field. The subdominant quadrupolar oscillations are stochastically forced by dominant dipolar oscillations via the equator-symmetric part of the fluctuating $\alpha$-effect. The amplitude and sense of the asymmetry of individual cycles varies on a time scale of the order of four dynamo-cycle periods. The variations are irregular, i.e. not periodic. The model suggests that asymmetry in the polar magnetic fields in the solar minima can be used as a precursor for asymmetry of sunspot activity in the following solar cycle.}
{}


    \titlerunning{Long-term hemispheric asymmetry of solar activity}
    \authorrunning{Alexander Nepomnyashchikh et al.}

    \maketitle
\section{Introduction}
The uneven distribution of sunspots and other manifestations of magnetic activity about the solar equator has long been known \citep[cf.][]{SM890,M904}. The origin of the hemispheric asymmetry is still a matter of debate \citep[see the recent publications by][and references therein]{NCP14,H15,Dea16,SK17,SC18}. This may be because the origin differs between short- and long-term asymmetry. The asymmetry of sunspot activity on the time scales of from one solar rotation period to one year comprises a substantial randomness. This is evident from multiple and seemingly irregular sign reversals of monthly-averaged asymmetry $N-S$ \citep[cf. Fig.1 in][]{BO17}; $N$ and $S$ are some activity measure for the northern and southern hemispheres respectively. Relative asymmetry $(N-S)/(N+S)$ is typically enhanced at the solar cycle minima \citep{JJ04,Mea17} while absolute asymmetry peaks at activity maxima \citep{Tea06}.
A similar trend can be expected for random events occurring with equal probability in either hemisphere, for which case $\mid N_N - N_S\mid\, \approx \sqrt{N_N + N_S}$ ($N_N$ and $N_S$ being the event numbers).

Solar cycle averaged hemispheric asymmetry, on the contrary, shows coherence over several activity cycles. \citet{OB94} found a wave-type secular trend in the asymmetry of sunspot areas for the epoch of 1874-1989 with southern and northern hemispheres being alternately more active for several solar cycles. \citet{BO11} confirmed this trend and extended it to the 1821-2009 epoch. \citet{Zea09} and \citet{Mea13} found that the activity cycles in the northern and southern hemispheres keep their phase difference for several cycles with the two hemispheres leading the variations alternately.

An explanation for the long-term coherence of the hemispheric asymmetry observed can be found in the dynamo theory. The theory allows two types of global magnetic modes of dipolar (anti-symmetric) and quadrupolar (symmetric) equatorial parity \citep{KR80}. Both modes are symmetric in the sense of magnetic energy (to which the observed magnetic activity is probably related). The asymmetry can result from a superposition of dipolar and quadrupolar modes. \citet{SC18} clearly explained the emergence of hemispheric asymmetry from a superposition of oscillating dipolar and quadrupolar dynamo modes. The doubly periodic asymmetry oscillates with beat and sum frequencies of the two periodic modes.

Solar dynamo models allow mixed-parity solutions if they include sufficiently strong nonlinearities \citep[cf.][]{SN94,T97,WT16} or if the model parameters lack a certain equatorial symmetry. The asymmetry in dynamo parameters is usually associated with their random fluctuations \citep[cf.][]{USM09,SC18,KMB18}. Observational gyrochronology of solar-type stars shows that the solar dynamo is only slightly supercritical and therefore weakly nonlinear \citep{vSea16}. The dynamo model of this paper is, therefore, weakly nonlinear but includes fluctuations in its key parameter.

The observed long-term coherence in the solar hemispheric asymmetry does not exclude its stochastic origin. Stochastically forced oscillations can keep coherence over many oscillation periods. $p$-modes of global solar oscillations give a relevant example. The oscillations are forced stochastically by turbulent convection but have quality factors in the order of one thousand \citep{GMK94}. An example from dynamo theory can be the generation of a large-scale coherent magnetic field by temporally and spatially incoherent helicity fluctuations \citep{VB97}.

This paper studies long-term hemispheric asymmetry induced by short-term fluctuations in the solar dynamo. For this purpose, a mean-field $\alpha\Omega$-dynamo model with stochastic fluctuations in the $\alpha$-effect is applied. The amplitude and correlation time of the fluctuations have been formerly constrained from sunspot statistics \citep{OCK13,KMN18}. In this paper, the sunspot data are used to further constrain the latitudinal profile of the fluctuations. The weakly nonlinear dynamo model produces fields of pure dipolar parity when the alpha-effect is regular or includes random fluctuations in time only. Allowance for random variations with latitude violates the equatorial symmetry. Quasi-periodic quadrupolar oscillations randomly forced by the dominant dipolar oscillations via the equator-symmetric fluctuations in the $\alpha$-effect are identified as the physical mechanism for the hemispheric asymmetry.
The computed statistics of 4000 magnetic cycles is analysed to estimate the amplitude and degree of coherence of the hemispheric asymmetry in neighboring magnetic cycles.

Observational inferences for design of the dynamo model are discussed in the next Sect.\,\ref{data}. Section\,\ref{model} describes the dynamo model and the method of allowance for fluctuations in its $\alpha$-effect. Section\,\ref{results} presents and discusses the results. The final Sect.\,\ref{conclusions} summarises our conclusions.
\section{Implications of sunspot data}\label{data}
The $\alpha$-effect of our model corresponds to its particular version, known as the Babcock-Leighton mechanism for poloidal magnetic field generation \citep{B61}. This mechanism is related to the empirical Joy's law \citep{Hea19} for the tilt angle between the line connecting centers of opposite polarities of a spot group and the local line of latitude. Fluctuations in this type of $\alpha$-effect are related to the scatter in spot areas and distances between the centers of opposite polarities among sunspot groups but primarily to the scatter in the tilt angles \citep{OCK13,JCS14}.

\begin{figure}
   \includegraphics[width=\hsize]{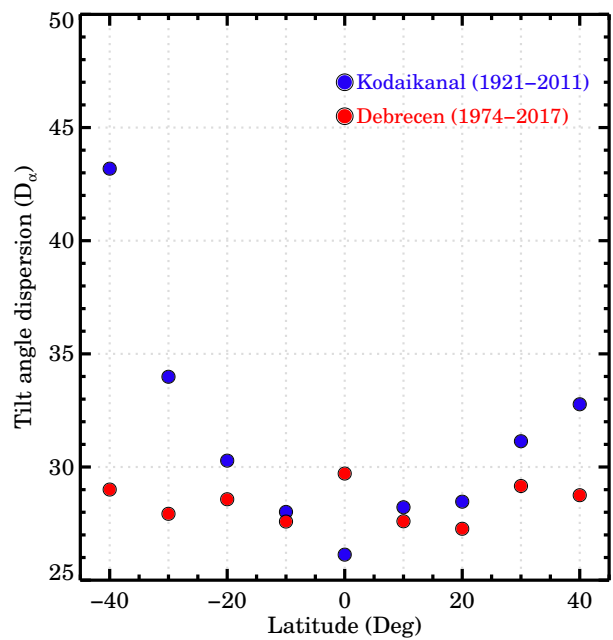}
   \caption{Dispersion of the tilt angles of Eq.\,(\ref{1}) estimated for the
            $10^\circ$ ranges of solar latitude. Circles with blue and red filling represent the sunspot data from Kodaikanal and Debrecen Observatory respectively. The blue point at $-45^\circ$ is severely affected by the poor statistics in that bin.
            }
   \label{f1}
\end{figure}

To evaluate the tilt-angle scatter, we used the sunspot area and position catalogue\footnote{\url{https://kso.iiap.res.in/new}} from the newly digitized Kodaikanal data \citep{Rea13,Mea17} and from the Debrecen sunspot data\footnote{\url{http://fenyi.solarobs.csfk.mta.hu/DPD/}} \citep{B15}. The method of tilt evaluation is the same as in \citet{H91}. Figure\,\ref{f1} shows the dispersion
\begin{equation}
    D_\alpha = \langle\left(\alpha - \langle\alpha\rangle\right)^2\rangle^{1/2}
    \label{1}
\end{equation}
of the tilt angle estimated for the $10^\circ$ ranges of solar latitude $\lambda = (-5^\circ\div 5^\circ), \pm(5^\circ\div15^\circ),...,\pm(35^\circ\div45^\circ)$; the angular brackets in this equation signify averaging over the ensemble of spot groups falling in a given range of latitude. The tilt-angle scatter of Fig.\,\ref{f1} varies moderately with latitude and it is not small in the equatorial region \citep[see also Table 2 in][]{JCS14}. This implies that the fluctuating part of the $\alpha$-effect does not fall to zero at the equator as the mean $\alpha$ value does.

\begin{table}
    \caption{Hemispheric Asymmetry of Spot Group Numbers for Solar Cycles 12--23}
    \label{t1}
    \centering
    \begin{tabular}{c c c c c}
        \hline
        Cycle No & $N_\mathrm{N}$ & $N_\mathrm{S}$ &
        $A_\mathrm{obs}$ & $P(|A_\mathrm{b}| \geq |A_\mathrm{obs}|)$ \\
        \hline
        12 & 646 & 1056 & -0.241 & 2.23$\times$10$^{-23}$ \\
        13 & 910 & 881  & 0.016  & 0.508 \\
        14 & 800 & 845  & -0.027 & 0.278  \\
        15 & 1172 & 1026 & 0.066 & 1.98$\times$10$^{-3}$ \\
        16 & 1042 & 932 & 0.056 & 1.41$\times$10$^{-2}$ \\
        17 & 1278 & 1248 & 0.012 & 0.564 \\
        18 & 1411 & 1419 & -0.003 & 0.895 \\
        19 & 1975 & 1443 & 0.156 & 8.97$\times$10$^{-20}$ \\
        20 & 1564 & 1320 & 0.085 & 5.97$\times$10$^{-6}$ \\
        21 & 1504 & 1545 & -0.013 & 0.469 \\
        22 & 1054 & 1286 & -0.099 & 1.76$\times$10$^{-6}$ \\
        23 & 1232 & 1409 & -0.067 & 6.13$\times$10$^{-4}$ \\
    \hline
    \end{tabular}
\end{table}

Next, we explore the hemispheric asymmetry in the sunspot group numbers. Sunspot statistics suit well for estimating the significance of this asymmetry by confronting it with a binomial random events model. The model assumes each event to occur with equal probability (=1/2) in the northern or southern hemisphere independently of other events. The probability $P(|A_\mathrm{b}| \geq |A_\mathrm{obs}|)$ of hemispheric asymmetry $A_\mathrm{b}$ of binomial random events to be equal or larger in absolute value than the observed asymmetry $A_\mathrm{obs} = (N_\mathrm{N} - N_\mathrm{S})/(N_\mathrm{N} + N_\mathrm{S})$, if sufficiently low, evidences the statistical significance of the observed asymmetry; $N_\mathrm{N}$ and $N_\mathrm{S}$ are the numbers of observed events in the northern and southern hemispheres respectively. The probability $P(|A_\mathrm{b}| \geq |A_\mathrm{obs}|)$ should, of course, be estimated for the same number of events $N = N_\mathrm{N} + N_\mathrm{S}$ as observed. The probability reads
\begin{eqnarray}
    P(|A_\mathrm{b}| \geq |A_\mathrm{obs}|) &=&
    \sum\limits_{n = 0}^{N_\mathrm{min}}
    \frac{N!}{2^{N-1}n!(N-n)!} ,
    \nonumber\\
    N_\mathrm{min} &=& \mathrm{min}(N_\mathrm{N},N_\mathrm{S}) .
    \label{2}
\end{eqnarray}
This equation applies to the case of $N_\mathrm{N} \neq N_\mathrm{S}$ (otherwise, the sought for probability equals one).

The binomial test applies only to countable objects, which the sunspots are. Spots are known to appear in groups. It makes sense, therefore, to apply the test to sunspot group numbers. Table\,\ref{t1} lists the numbers of sunspot groups, as counted from RGO\footnote{\url{https://solarscience.msfc.nasa.gov/greenwch.shtml}} (1874-1976) and Kislovodsk\footnote{\url{http://158.250.29.123:8000/web/Soln_Dann/}} (1977-2009) observatories, for solar cycles 12 to 23 separately for the two hemispheres. The Table shows also the corresponding relative hemispheric asymmetry and the probability of reproducing the asymmetry with the binomial random events model.

The table shows that the asymmetry in spot group number, if not too small (\lower.4ex\hbox{$\;\buildrel >\over{\scriptstyle\sim}\;$}$0.03$), is statistically significant. Cycles 12 and 22, 23 had significant asymmetry of the southern type. Cycles from 13 to 21 however had either insignificant asymmetry or clear predominance of the northern hemisphere. Hence, the asymmetry of a certain sense keeps coherence over several solar cycles.
\section{Dynamo model}\label{model}
Our dynamo model belongs to the so-called flux-transport models first developed by \citet{D95} and \citet{CSD95}. This name reflects the significance of meridional flow for latitudinal migration of magnetic fields. The flux-transport models with the $\alpha$-effect of Babcock-Leighton type show close agreement with solar observations \citep{Jea13}.

The model of this paper is very close to the model described in detail in former publications \citep{KN17MNRAS,KN17AL}. The only difference is in the random variations of the $\alpha$-effect in latitude and time now involved. We therefore describe only the method of accounting for these random variations in all necessary details but scantily outline the model design.
\subsection{Model design}
Our mean-field dynamo model solves numerically the axisymmetric 2D dynamo equations in the spherical shell of the convection zone. Performance of the model was tested and confirmed by comparison with the dynamo benchmark of \citet{Jea08}.

The dynamo model borrows the differential rotation and meridional flow from the differential rotation model of \citet{KO11}. The angular velocity profile is close to the results of helioseismology. The one-cell meridional flow is similar to the recent seismological inversions by \citet{RA15} and \citet{Mea18} which are distinct from other seismological detections of the meridional circulation in their satisfying the mass conservation constraint.

The differential rotation model also supplies the mean entropy gradient in terms of which the turbulent magnetic diffusivity is specified. The diffusivity varies smoothly about the value of $3\times 10^{12}$\,cm$^2$s$^{-1}$ in the bulk of the convection zone but drops by almost four orders of magnitude near the base of this zone. The decrease in the diffusivity near the bottom boundary produces the downward diamagnetic pumping effect which is important for the performance of solar dynamo models \citep{KKT06,GG08}. The downward pumping concentrates magnetic fields in the near-bottom region of low diffusion and provides a combination of relatively weak ($\sim10$\,G) polar fields with about one thousand times stronger near-bottom toroidal fields.

The turbulent diffusion and diamagnetic pumping are anisotropic. This means that the diffusion coefficient for the direction along the rotation axis is larger than the diffusivity for the direction normal to this axis and the diamagnetic pumping velocity depends on the magnetic field orientation relative to the rotation axis \citep{Oea02}. The anisotropy is induced by rotational influence on turbulent convection.
\subsection{Regular and fluctuating parts of the $\alpha$-effect}
In spherical geometry, magnetic fields $\vec B$ can be split in their toroidal and poloidal parts. In case of axial symmetry, this splitting reads
\begin{equation}
    \vec{B} = \vec{e}_\phi B(r,\theta) +
    \vec{\nabla}\times\left(\vec{e}_\phi\frac{A(r,\theta)}{r\sin\theta}\right)\ ,
    \label{3}
\end{equation}
where $B$ is the toroidal field, $A$ is the poloidal field potential, standard spherical coordinates are used ($\theta$ is the co-latitude), and $\vec{e}_\phi$ is the azimuthal unit vector. The $\alpha$-effect of mean-field MHD converts toroidal fields into poloidal fields and vice versa \citep[cf.][]{KR80}.

The poloidal field equation of the $\alpha\Omega$-dynamos reads
\begin{equation}
    \frac{\partial A}{\partial t} = \frac{1}{\rho r\sin\theta}
    \left(\frac{\partial\psi}{\partial r}\frac{\partial A}{\partial\theta} +
    \frac{\partial\psi}{\partial\theta}\frac{\partial A}{\partial r}\right) +
    r\sin\theta\ {\cal E}_\phi ,
    \label{4}
\end{equation}
where $\psi$ is the meridional flow stream function, $\rho$ is density, and ${\cal E}_\phi$ is the azimuthal component of the mean electromotive force, which includes the additive part ${\cal E}^\alpha_\phi$ responsible for the $\alpha$-effect \citep{KR80}. In our model, this contribution of the $\alpha$-effect is specified as follows,
\begin{equation}
    {\cal E}^\alpha_\phi = \alpha\ \frac{\sin^{n_\alpha}\theta\ \phi_\alpha(r)}
    {1 + (B(r_\mathrm{i},\theta)/B_0)^2} f(\theta,t)\ B(r_\mathrm{i},\theta).
    \label{5}
\end{equation}
The $\alpha$-effect of Eq.\,(\ref{5}) is non-local: it generates the poloidal field near the surface from the bottom toroidal field, $r_\mathrm{i}$ is the radius of the inner boundary. The non-local effect is supposed to represent the Babcock-Leighton mechanism. The relatively large value of $n_\alpha = 7$ reflects the spot emergence near the solar equator. The positive function $\phi_\alpha$ peaks near the top boundary \citep{KN17MNRAS}. The denominator in Eq.\,(\ref{5}) accounts for the magnetic quenching of the $\alpha$-effect ($B_0 = 10$\,kG). The coefficient $\alpha$ measures the intensity of the $\alpha$-effect. Its threshold value for the self-sustained dynamo in our model is $\alpha_\mathrm{t} = 0.158$\,m\,s$^{-1}$. Stellar gyrochronology suggests that the rotation rate of the Sun is only slightly above the threshold rate for the global dynamo \citep{MvS17}. The $\alpha$-parameter is estimated to be about 10\% supercritical \citep{KN17MNRAS}. Computations of this paper are therefore performed with $\alpha = 0.174$\,m\,s$^{-1}$.

The function
\begin{equation}
    f(\theta ,t) = \cos\theta + \frac{1}{4}\sigma S(\theta,t)
    \label{6}
\end{equation}
in Eq.\,(\ref{5}) includes regular and fluctuating parts by the first and second terms in its right-hand side respectively. In this equation, $\sigma$ is the relative amplitude of the fluctuations, $S(\theta,t)$ is a random function of order one, and the factor $1/4$ is close to the value of $\cos\theta$ at the latitude $\lambda = 15^\circ$ where the near-bottom toroidal field of our model attains its largest values ($\lambda = \pi/2 - \theta$). The fluctuations of the $\alpha$-effect can be switched off by putting $\sigma = 0$ in Eq.\,(\ref{6}). The remaining regular part is antisymmetric about the equator as it should be for the effect originating from the Coriolis force. The tilt-angle dispersion of Fig.\,\ref{f1}, however, indicates that the fluctuating part of the $\alpha$-effect should not vanish at the equator (though the electromotive force (\ref{5}) may vanish if the bottom toroidal field does so).

Following \citep{OK13}, the random function $S(\theta, t)$ is modelled by solving the equations
\begin{eqnarray}
    \frac{\mathrm{d}S}{\mathrm{d}t} &=& -\frac{n}{\tau}\left( S - S_1\right)\ ,
    \nonumber \\
    \frac{\mathrm{d}S_1}{\mathrm{d}t} &=& -\frac{n}{\tau}\left( S_1 - S_2\right)\ ,
    \nonumber \\
     ... &&
    \nonumber \\
    \frac{\mathrm{d}S_{n-1}}{\mathrm{d}t} &=& -\frac{n}{\tau}\left( S_{n-1} - (2\pi)^{1/4}\sqrt{\frac{2\tau}{\delta t\ \delta\theta}}\ \hat{g}
    \exp\left(-\left(\frac{\theta - \theta_0}{\delta\theta}\right)^2\right)\right)
    \label{7}
\end{eqnarray}
in line with the dynamo equations. In these equations, $\tau$ is the correlation time and $\delta\theta$ is the correlation angular distance in latitude, $\delta t$ is the numerical time step, $\hat g$ is the normally distributed random number with zero mean and {\sl rms} value equal one, and $\theta_0$ is the random colatitude distributed uniformly in the range of $[0,\pi]$. The values of $\hat g$ and $\theta_0$ were renovated on each time step independently of their previous values. Computations of this paper were done with $n = 3$ in Eqs.\,(\ref{7}).

\begin{figure}
   \includegraphics[width=7 truecm]{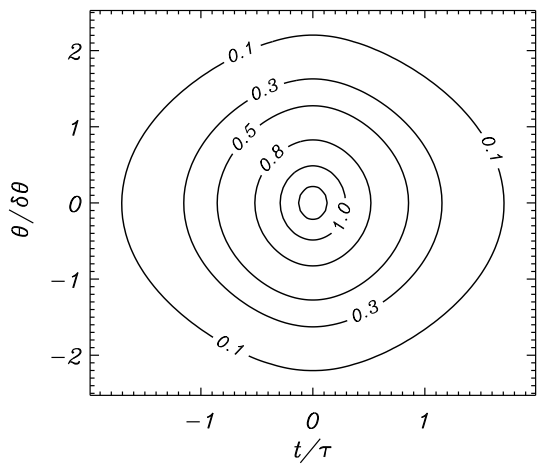}
   \caption{Correlation function (\ref{8}) of normalised fluctuations
            of the $\alpha$-effect.
            }
   \label{f2}
\end{figure}

The random process of Eqs.\,(\ref{7}) can be characterised by the correlation function
\begin{equation}
    C(\theta,t) = \langle S(t',\theta')S(t' + t,\theta' + \theta)\rangle ,
    \label{8}
\end{equation}
where angular brackets signify averaging over time $t'$. Provided that $\theta'$ in Eq.\,(\ref{8}) is not too close to the poles, $\delta\theta < \theta' < \pi - \delta\theta$, the correlation function depends on the time $t$ and latitudinal distance $\theta$ through the ratios $t/\tau$ and $\theta/\delta\theta$ only. The numerically defined correlation function is shown in Fig.\,\ref{f2}. The characteristic appearance of the random function $S(\theta,t)$ can be seen in fig.\,4 of \citet{OK13}.

The values of $\sigma = 2.7$ and $\tau = 25.4$\,days are used in the computations of this paper. These values are inferred from sunspot statistics \citep{OCK13,KMN18}. The value of $\delta\theta = 0.1$ ($\simeq 6^\circ$) is close to the characteristic latitudinal size of sunspot groups.
\subsection{Equator-symmetric part of the $\alpha$-effect as a source of hemispheric asymmetry}\label{symmalpha}
Magnetic fields of the dynamo model can be thought of as a superposition of dipolar field combining the equator-antisymmetric part of the toroidal field $B$ with the equator-symmetric potential $A$,
\begin{eqnarray}
    B_\mathrm{d}(\theta) &=& (B(\theta) - B(\pi-\theta))/2,
    \nonumber \\
    A_\mathrm{d}(\theta) &=& (A(\theta) + A(\pi-\theta))/2,
    \label{9}
\end{eqnarray}
and the quadrupolar field
\begin{eqnarray}
    B_\mathrm{q}(\theta) &=& (B(\theta) + B(\pi-\theta))/2,
    \nonumber \\
    A_\mathrm{q}(\theta) &=& (A(\theta) - A(\pi-\theta))/2.
    \label{10}
\end{eqnarray}
If the fluctuations are omitted ($\sigma = 0$ in Eq.\,({6})), the dynamo can converge to a definite parity. Our model eventually converges to a dipolar field from an initial field of mixed parity. The dynamo field does not deviate from the equator-symmetric configuration only if the initial field is purely quadrupolar. This subdominant quadrupolar solution oscillates with an (energy) cycle period of 10.4 yr, which is close to the period of 10.7 yr of the dominant dipolar mode. Variable hemispheric asymmetry cannot, therefore, result in our model from beat phenomenon of different parity modes.
\begin{figure}
   \includegraphics[width=\hsize]{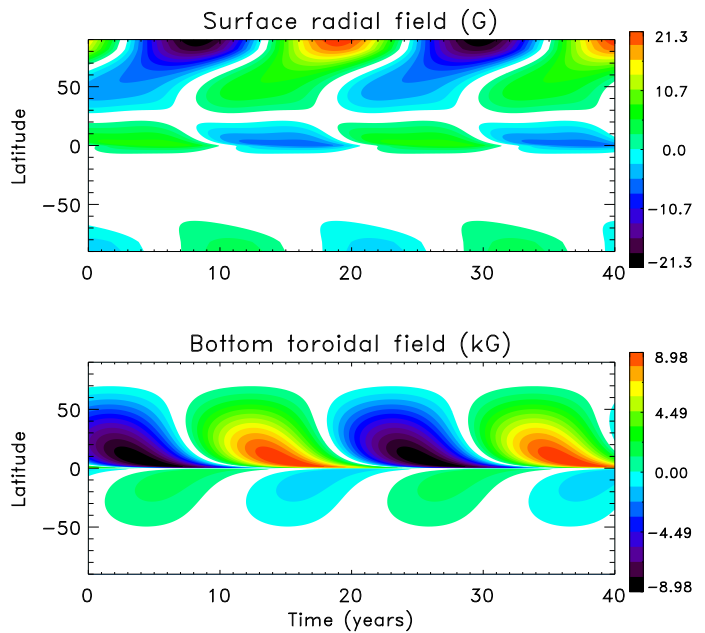}
   \caption{Time-latitude pattern of radial field on the surface and toroidal
            field at the bottom boundary computed with the equator-asymmetric profile            (\ref{11}) of the alpha-effect.
            }
   \label{f3}
\end{figure}

The linear marginal value $\alpha^q_c = 0.169$\,m\,s$^{-1}$ for excitation of quadrupolar modes is larger than $\alpha^d_c = 0.158$\,m\,s$^{-1}$ for dipolar modes but slightly smaller than $\alpha = 0.174$\,m\,s$^{-1}$ used in our computations. Nevertheless, the computations converged to dipolar parity whenever the $\alpha$-effect was antisymmetric about the equator. This is probably because the weak nonlinearity of Eq.\,(\ref{5}) reduces the effective value of $\alpha$ below the $\alpha^q_c$.

If the equator-symmetric part of the random fluctuations in the $\alpha$-effect of Eq.\,(\ref{5}) is filtered out, the dynamo still converges to the dipolar field.
It can be seen from Eqs.\,(\ref{4}) and (\ref{5}) that the presence of an equator symmetric part in the $\alpha$-effect couples the dipolar and quadrupolar fields so that solutions of pure parity are no longer possible.

Figure\,\ref{f3} shows the solution computed with the Eq.\,(\ref{6}) replaced by the profile
\begin{equation}
    f(\theta) = \cos\theta + 0.1
    \label{11}
\end{equation}
containing a small positive equator-symmetric part. The profile is enhanced in the northern hemisphere. Figure\,\ref{f3} shows accordingly the asymmetry of northern type. If the sign of the symmetric part of the profile (\ref{11}) is reversed to $f(\theta) = \cos\theta - 0.1$, the pattern of Fig.\,\ref{f3} changes to its mirror-reflection about the equator.

Fluctuations in the $\alpha$-effect containing equator-symmetric part result in hemispheric differences in the phase, duration and amplitude of magnetic cycles
\citep{USM09,OK13,Pea14,SC18,KMB18}. It is therefore uncertain how the instants of cycle beginning, maxim and termination should be defined. However, the amplitude of dipolar part (\ref{9}) of the magnetic field considerably exceeds the quadrupolar part in \lq almost all' cycles of our model (about 0.3\% of computed cycles only had a quadrupol predominance). We therefore defined the end of a magnetic cycle and commencement of the next cycle as the instant of reversal of the dipolar part $B_\mathrm{d}$ of the bottom toroidal field at the latitude of $15^\circ$ where the toroidal field attains its largest strengths. Accordingly, the cycles maxima are defined as the instants of the strongest $B_\mathrm{d}$ in a given cycle.

Hemispheric asymmetry was evaluated in our computations by its relative value
\begin{equation}
    A_\mathrm{t} = \frac{B^2(15^\circ) - B^2(-15^\circ)}{B^2(15^\circ) + B^2(-15^\circ)}
    \label{12}
\end{equation}
for the bottom toroidal field of the cycles maxima (the argument of the toroidal field in this equation is the latitude $\lambda = \pi/2 - \theta$). Another measure for the asymmetry,
\begin{equation}
    A_\mathrm{p} = \frac{B_r^2(90^\circ) - B_r^2(-90^\circ)}{B^2(90^\circ) + B_r^2(-90^\circ)}
    \label{13}
\end{equation}
is estimated with the surface polar field ($B_r$) for the cycles' minima. The importance of polar fields of the cycles minima is related to their tight correlation with the amplitude of the following cycles \citep{Sea78,CCJ07,WS09}. Another measure of equatorial parity traditionally used in dynamo modeling is related to magnetic energy ($E = E_\mathrm{d} + E_\mathrm{q}$) composed of the energy of the dipolar ($E_\mathrm{d}$) and quadrupolar ($E_\mathrm{q}$) field components:
\begin{equation}
    P = \left(E_\mathrm{q} - E_\mathrm{d}\right)E^{-1}.
    \label{14}
\end{equation}
The parity value at the cycles maxima and the time-averaged parity of individual cycles were included into the statistics of magnetic cycles of our computations together with the asymmetry parameters of Eqs.\,(\ref{12}) and (\ref{13}).

The equator-symmetric part of the random function $S(\theta,t)$ in Eq.\,(\ref{6}) varies on the short time scale of solar rotation and it is zero on average. Only computations can show whether the short-term fluctuations can produce the hemispheric asymmetry coherent over several activity cycles as observed on the Sun.
\section{Results and discussion}\label{results}
Results of this Section refer to the statistics of 4000 computed cycles. Hemispheric asymmetry parameters averaged over the entire ensemble of computed cycles, $\langle A_\mathrm{t}\rangle = 3.5\times 10^{-3},\ \langle A_\mathrm{p}\rangle = 3.7\times 10^{-3}$, are very small. Individual cycles can be considerably asymmetric however.

\begin{figure}
   \includegraphics[width=\hsize]{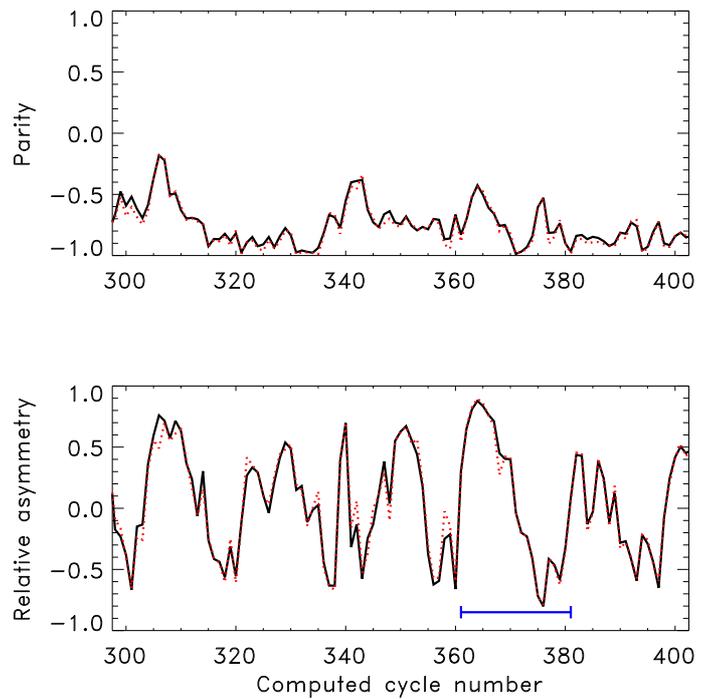}
   \caption{{\sl Top panel:} time-averaged parity of individual cycles
            (full line) and instant values of the parity (\ref{14}) at the cycles maxima (dotted) as the functions of computed cycle number. The data points for individual cycles are connected by straight lines for
            better visibility. {\sl Bottom panel:} The asymmetry measure (\ref{12}) of the toroidal field for the cycles maxima (full line) and the asymmetry (\ref{13}) of the polar fields at the cycle beginnings (dotted). The straight horizontal line indicates the range, for which the next two plots show the detailed time dependencies.
            }
   \label{f4}
\end{figure}

Figure\,\ref{f4} shows a characteristic fragment of the asymmetry parameters for the computed cycles 300$\div$400. The cycle-averaged parity and its instantaneous values at the cycles maxima are close together. Therefore, the parity varies little in the course of individual cycles. The negative parity shows the predominance of dipolar fields. The polar field asymmetry at the cycle beginnings correlates closely with the toroidal field asymmetry at the cycle maxima. The amplitudes of solar cycles are known to be predetermined by the polar field strength of the preceding activity minima \citep[cf.][]{Sea78,CCJ07,WS09}. The model computations suggest that the cycles asymmetry is also controlled by the poloidal field structure of preceding minima.

\begin{figure}
   \includegraphics[width=\hsize]{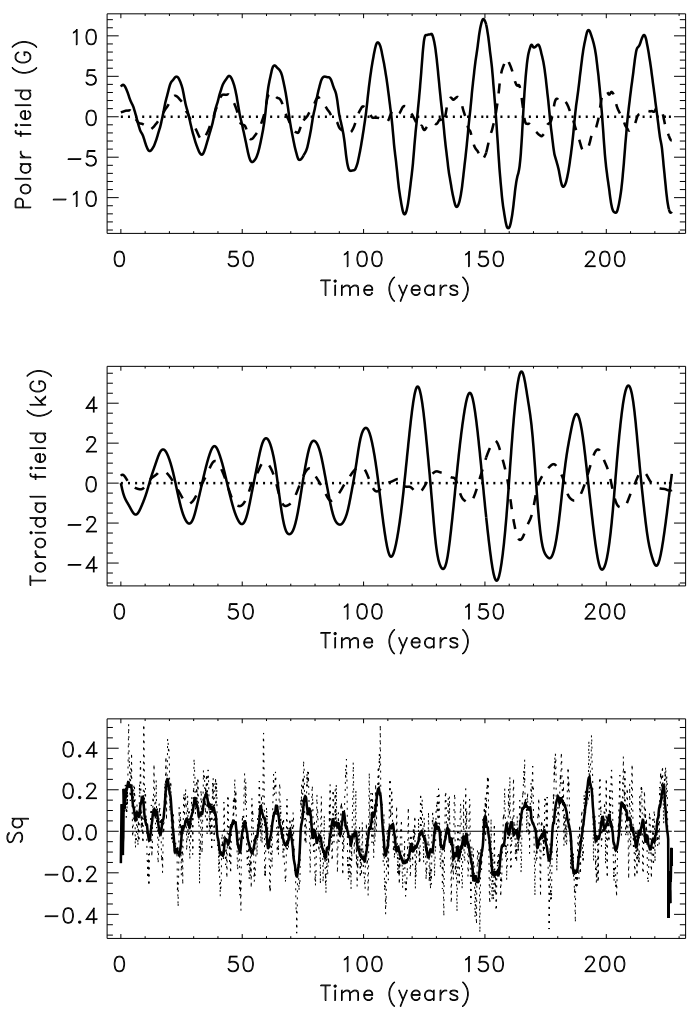}
   \caption{{\sl Top panel:} full line shows the equator-antisymmetric
            (dipolar) part of the polar field for the computed cycles 361$\div$381. The equator-symmetric (quadrupolar) part of the field is shown by the dashed line. {\sl Middle panel:} the same as in the top panel but for the bottom toroidal field at the latitude of $15^\circ$. {\sl Bottom panel:} normalized equator-symmetric part of the $\alpha$-fluctuations of Eq.\,(\ref{15}) (dashed line) and its three-years running mean (full line).
            }
   \label{f5}
\end{figure}

\begin{figure}
   \includegraphics[width=\hsize]{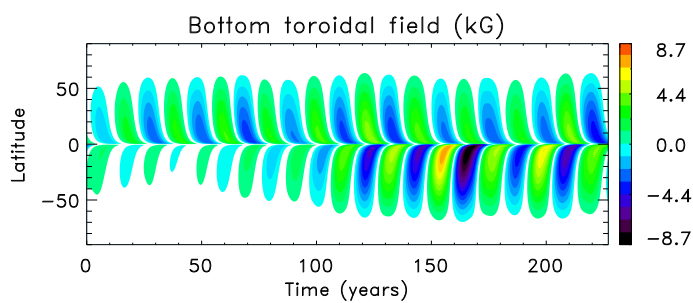}
   \caption{Time-latitude diagram of the bottom toroidal field for the same
            computation as Fig.\,\ref{f5}.
            }
   \label{f6}
\end{figure}

Figure\,\ref{f4} shows also that neighboring cycles tend to have asymmetry of the same sense. The computed cycles 361$\div$381 indicated by a straight horizontal line in Fig.\,\ref{f4} show the northern type asymmetry in the first ten cycles of the group and a change to southern type asymmetry in the last ten cycles. Figures\,\ref{f5} and \ref{f6} show the variations of the field components with time for this group of cycles. Figure\,\ref{f5} shows also the latitude-averaged symmetric part $S_\mathrm{q}$ of the fluctuating $\alpha$-effect
\begin{equation}
    S_\mathrm{q}(t) = \int\limits_0^{1/2}
    \left(S(\theta,t) + S(\pi - \theta,t)\right) \mathrm{d}(\cos\theta).
    \label{15}
\end{equation}
The averaging is over the near-equatorial region where the alpha-effect of Eq.\,(\ref{5}) operates.

Fluctuations of the $\alpha$-effect vary on a short time scale and vanish on average. The bottom panel of Fig.\,\ref{f5}, nevertheless, shows small but finite fluctuations left after three-years time averaging. The equator-symmetric fluctuations were shown in Sect.\,\ref{symmalpha} to result in the hemispheric asymmetry of dynamo-generated fields. At the beginning of the run of Fig.\,\ref{f5}, the fluctuations were predominantly positive. This results in the quadrupolar part of the field synchronised in phase with the dipolar oscillations. Accordingly, the northern hemisphere was dominant initially (Fig.\,\ref{f6}). In the middle of the run, the $S_\mathrm{q}$ parameter changed to predominantly negative. This is the probable reason for the change in the dipolar and quadrupolar parts of the field to anti-phase oscillations and predominance of the southern hemisphere seen at the end of the run in  Fig.\,\ref{f6}.

\begin{figure}
   \includegraphics[width=\hsize]{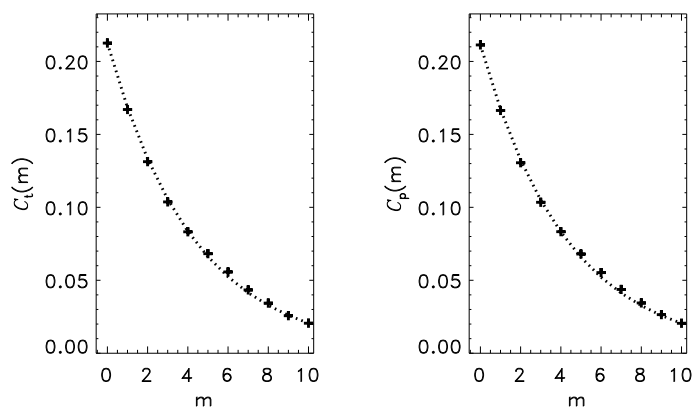}
   \caption{The asymmetry correlation functions of Eqs.\,(\ref{16}) and (\ref{17}).
            The dotted lines show the approximation by the exponential law $\exp(-m/m_\mathrm{c})$ ($m_\mathrm{c} = 4.28$).
            }
   \label{f7}
\end{figure}

The asymmetry of Fig.\,\ref{f4} may give an (illusive) impression of its long-term periodicity. The long-term coherence of the asymmetry and its possible periodicity can be probed by computing the correlation functions
\begin{eqnarray}
    C_\mathrm{t}(m) &=& \frac{1}{N-m}\sum\limits_{n=1}^{N-m}
    A_\mathrm{t}(n)A_\mathrm{t}(n+m) ,
    \label{16} \\
    C_\mathrm{p}(m) &=& \frac{1}{N-m}\sum\limits_{n=1}^{N-m}
    A_\mathrm{p}(n)A_\mathrm{p}(n+m) ,
    \label{17}
\end{eqnarray}
where $A_\mathrm{t}(n)$ and $A_\mathrm{p}(n)$ are the asymmetries of the toroidal and poloidal field of Eqs.\,(\ref{12}) and (\ref{13}) respectively in the $n$-th computed cycle, and $N$ is the total cycle number ($4000$ in our case).

The correlation functions are shown in Fig.\,\ref{f7}. The functions are closely approximated by the exponential law $\exp(-m/m_\mathrm{c})$. The coherence number $m_\mathrm{c} = 4.28$ means that the sense of hemispheric asymmetry is kept for on average about 4 successive cycles. A periodicity in the asymmetry variations would result in a change of the correlations to negative values with increasing $m$. The absence of such a sign reversal in Fig.\,\ref{f7} means that the asymmetry of our model is not periodic.

Linear oscillation frequencies of dipolar and quadrupolar modes differ little in our model. According to \citet{SC18}, this implies a small beat frequency and doubled sum frequency for the asymmetry oscillations. In other words, the asymmetry of a given sense should persist for many cycles and the coherence number $m_\mathrm{c}$ should be large. The reason for the moderate $m_\mathrm{c}$ of our model probably is that its magnetic cycles are not periodic. The cycle durations and amplitudes vary randomly from cycle to cycle \citep{KMN18}. The beat phenomenon does not apply to the quasi-periodic oscillations. Random wandering of phase shifts between the dipolar and quadrupolar oscillations (Figs.\,\ref{f5} and \ref{f6}) reduces the coherence number to about four magnetic cycles.

\begin{figure}
   \includegraphics[width=8 truecm]{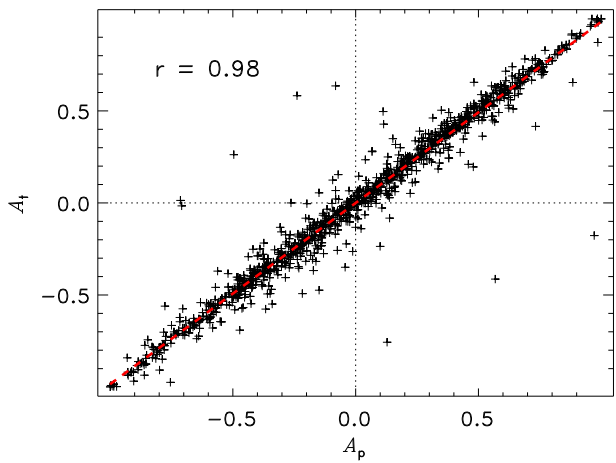}
   \caption{Scatter plot of the computed cycles on the plane of
            the toroidal field asymmetry of Eq.\,(\ref{12}) of the cycles maxima and the polar field asymmetry (\ref{13}) of preceding minima. The dashed line shows the best linear fit. Each fourth of the computed cycles only is shown to avoid jam in the plot.
            }
   \label{f8}
\end{figure}

The similarity of the poloidal and toroidal field asymmetries in Figs.\,\ref{4} and \ref{f7} is indicative of their correlation. Figure\,\ref{f8} confirms that the poloidal field asymmetry of the cycles minima and toroidal field asymmetry of the following maxima are tightly correlated. The correlation coefficient for the plot of Fig.\,\ref{f8} is $r = 0.98$. The model predicted correlation if confirmed observationally can be of certain predictive significance. The solar cycle amplitudes can be predicted from the measured strengths of the large-scale polar fields of the preceding minima \citep[cf.][and references therein]{CCJ07,HU16}. The correlation of Fig.\,\ref{f8} suggests that the hemispheric asymmetry can also be predicted from the same measurements.
\section{Conclusions}\label{conclusions}
The scatter in the the tilt angles of the sunspot groups observed shows that the fluctuating part of the Babcock-Leighton type $\alpha$-effect of the solar dynamo does not vanish at the equator.

The accordingly-designed dynamo model shows the long-term hemispheric asymmetry of the simulated magnetic cycles resulting from short-term fluctuations in the $\alpha$-effect. The physical mechanism for this asymmetry is the excitation of the subdominant quadrupolar dynamo mode by the dominant dipolar mode via the equator-symmetric part of the fluctuating $\alpha$-effect.

Statistical analysis of the computed magnetic cycles shows that the sense and amplitude of hemispheric asymmetry varies irregularly on a characteristic time scale of several (about four) cycles. The variations are non-periodic.

Statistical analysis of dynamo computations suggests that the asymmetry of solar activity cycles can be predicted from the asymmetry of the polar magnetic field of the preceding activity minima.
\begin{acknowledgements}
The authors are thankful to A.\,A.~Osipova (Pulkovo Observatory) for preparing sunspot data for Table\,\ref{t1}. AN and LK acknowledge the support by the Russian Foundation for Basic Research (project 17-52-80064\_BRICS) and by budgetary funding of Basic Research program II.16. DB acknowledges the support from DST, India for BRICS project grant.
\end{acknowledgements}
\bibliographystyle{aa}
\bibliography{paper}
\end{document}